\documentclass[12pt]{article}
\usepackage{amssymb}
\usepackage{amsmath}
\usepackage{graphicx}
\usepackage[matrix,arrow]{xy}
\parskip=\medskipamount
\def\om{\omega}   
\def\R{{\bf R}}   
\def\IC{\relax{\it  l\kern-.50 em C}}
\def\IE{\relax{\it  l\kern-.12 em E}}
\def\IK{\relax{\it  l\kern-.18 em K}}
\def\IL{\relax{\it  I\kern-.18 em L}}
\def\IN{\relax{\it  I\kern-.18 em N}}
\def\IR{\relax{\it  I\kern-.18 em R}}

\def\<#1>{\langle#1\rangle}
\def\d<#1>{\langle\langle#1\rangle\rangle}
\font\tenfrak=eufm10  \font\sevenfrak=eufm7  \font\fivefrak=eufm5
\newfam\frakfam
\textfont\frakfam=\tenfrak
\scriptfont\frakfam=\sevenfrak
\scriptscriptfont\frakfam=\fivefrak

\newtheorem{teorema}{Theorem}

\def\smallonehalf{\frac{{}_1}{{}^2}}

\def\frac#1#2{{#1\over #2}}

\def\Cos{\mathop{\rm C}\nolimits}    
\def\Sin{\mathop{\rm S}\nolimits}    
\def\k{\kappa}                       

\def\pd#1#2{\frac{\partial #1}{\partial#2}}

\def\Hil{\mathcal{H}}
\def\SH{S\mathcal{H}}
\def\PH{P\mathcal{H}}
\begin{document}

\title{A geometric approach to a generalized virial theorem}

\author{
Jos\'e F. Cari\~nena$^{a)}$, Fernando Falceto$^{b)}$ and Manuel F.
Ra\~nada$^{c)}$
\\[4pt]
  {\sl Departamento de F\'{\i}sica Te\'orica, Facultad de Ciencias} \\
  {\sl Universidad de Zaragoza, 50009 Zaragoza, Spain}
}
\maketitle

\begin{abstract}
The virial theorem, introduced by Clausius in statistical
mechanics, and later  applied in both classical mechanics and
quantum mechanics, is studied by making use of  symplectic
formalism as an approach  in the case of both  the Hamiltonian and Lagrangian
systems. The possibility of establishing virial's like  theorems from  one-parameter groups of non-strictly canonical transformations is analysed;
and the case of systems with a position dependent mass is also
discussed. Using the modern symplectic approach to quantum mechanics we arrive at
the quantum virial theorem in full analogy with the classical case.  
\end{abstract}
\begin{quote}

{\sl Running title:}{\enskip}
A geometric approach to a generalized virial theorem.




\end{quote}
{\vfill}

\footnoterule
{\noindent\small
$^{a)}${\it E-mail address:} {jfc@unizar.es} \\
$^{b)}${\it E-mail address:} {falceto@unizar.es}\\
$^{c)}${\it E-mail address:} {mfran@unizar.es}  }
\newpage

\section{Introduction }
\indent

In 1870 (see \cite{Collins} for an historical
account), Clausius introduced  the  virial function for a one particle system
\begin{equation}
 G({\bf r},  {\bf v})=m\, {\bf r}\cdot {\bf v}\label{clausius}
\end{equation}
for studying  the motion of   a  particle of  mass $m$ under
the action of a force ${\bf F}$. Then, using  such a function and
the Newton second law, he proved that  when
either  the motion is  periodic of period $T$,
 or the motion
is not periodic but  the possible values of the function $G$ are
bounded  and we take the limit of $T$ going to infinity, the time average over a
time interval $T$ of   kinetic energy $E_c({\bf v})$
 is given by
\begin{equation}
 \d<E_c({\bf v})> = - \frac{1}{2}\d< {\bf r}\cdot {\bf F}>,\label{sVT}
\end{equation}
so that, in the particular case of a  conservative force, he obtained
\begin{equation}
 \d<E_c({\bf v})> = \frac{1}{2}\d< {\bf r}\cdot \boldsymbol{\nabla}V> \,.\label{sVTcf}
\end{equation}
 If the potential  $V$ is
homogeneous of degree $k$, Euler's theorem of homogeneous
functions implies that ${\bf r}\cdot \boldsymbol{\nabla}V=k\, V$,
and therefore,
$
 2\,\d<E_c({\bf v})> = k\,\d<V({\bf r})>$,
that leads to the following values for the averages of the kinetic
energy and the potential
$$
 \d<E_c({\bf v})>  =  \frac{k\,E}{k+2} \,,{\qquad}
 \d<V({\bf r})>  = \frac{2\,E}{k+2}  \,,
$$
where $E$ is the total energy \cite{AvB}.  For instance in the
harmonic oscillator case,  with $k=2$,  and in the  Kepler problem
with negative energy, with $k=-1$, we  obtain, respectively,
$$
 \d<E_c({\bf v})> = \d<V({\bf r})> = \frac{1}{2} E \,,
$$
$$  \d<E_c({\bf v})> = -E  ,{\qquad }\d<V({\bf r})> = 2\,E \,.
$$

The virial theorem was originally introduced in classical
statistical mechanics  (that was the original matter  studied by
Clausius)  but later  became important in many other
different branches of physics. The quantum mechanical version of
the theorem is due to Born, Heisenberg, and Jordan \cite{BHJ26}
and presently it is a tool frequently used in  many body quantum
mechanics (e.g. systems of  fermions) and in molecular physics; in
addition,  this matter has  also been related to  certain
fundamental questions arising in nonrelativistic quantum mechanics
as the Hellmann--Feynman theorem (see  \cite{Lowd59}-\cite{BeN10}
and references therein).  We  also mention the importance of the
virial theorem  in solar and stellar astrophysics \cite{EvA05}.
The important point therefore is the  wide range of applicability
of the virial theorem with applications ranging from  dynamical
(even relativistic) and thermodynamical systems, to the dust and
gas of interstellar space, as well as cosmological  considerations
of the universe as a whole and in other  discussions concerning
the stability of clusters, galaxies and clusters of galaxies
\cite{Sa85}, or giving information on the masses of bound systems,
being the main reason we think that dark matter exists
\cite{AEv11, ChTV11}. Of course this virial theorem provides less
information that the equations of motion but it is simpler to
apply  and give us some information on  systems whose complete
analysis may defy description.

 After this standard approach to virial theorem several questions
appear. For instance,  the role played by  the function $G$  and
also the reason  why the relation is  simpler for power law
potentials. There are more general results known as hypervirial
theorems introduced in \cite{Hi60} whose natural framework is the
theory of locally Hamiltonian dynamical systems.  Moreover, quantum
mechanics is also a particular case of this more general theory
which suggests us to analyse the problem from this new perspective,
that then  enables us to develop the corresponding theorem in quantum
mechanics.  The main point is that the standard theory is always
related with scaling and dilation transformations
\cite{BrFr77,Kl79,Lowd59,vW78,YaW78}; there is no room for such
transformations when the configuration space is no longer a linear
space but there exist, for example,  holonomic constraints.

 Concerning the last question, the search for generalizations is a
matter that has been studied by different authors from different
viewpoints  (in fact the quantum mechanical virial for the
expectation values of a quantized system can be considered as a
generalization of the original virial theorem). Here we mention a
version for discrete maps \cite{How05} or a virial theorem in a
spherical geometry \cite{LiZhang11}. An interesting result is that
it has been  related with the Noether's theorem and with the
invariance properties of the Lagrange function
\cite{vKa72,NaVV86}.

 The main aim of this paper is to develop a deeper analysis of the virial
theorem, both in classical and in quantum mechanics, using  the
modern theory of Hamiltonian systems in symplectic manifolds as an
approach, and therefore including Lagrangian formulations.

The paper is organised as follows. In Section 2, a generalized
virial theorem is established in the framework  of  the theory of
Hamiltonian systems in symplectic manifolds.  In Section 3, the
virial  theorem is related with the Lagrangian formalism by
considering the Lagrangian systems as Hamiltonian systems in
tangent bundle manifolds; we prove that not only point symmetries of $L$
are useful in classical mechanics but more general transformations  modifying the Lagrangian by
a factor, can be used to establish virial type relations. Two
generalizations of the virial theorem are studied in Sections 4
and 5. In Section 4 we consider the case of systems with a
position dependent mass and in Section 5 we relate the theorem
with the theory of non-strictly canonical transformations. We
study in Section 6 the virial theorem in quantum mechanics using
 quantum symplectic formalism  as an approach. Finally, in
Sec. 7 we  make some comments and present some open questions.

\section{Virial theorem for Hamiltonian systems in symplectic manifolds}

To make the paper  self-contained  and to introduce the notation,  
we  recall some concepts of the geometrical approach to
the theory of  Hamiltonian systems.  Then we
consider the virial theorem from a geometric perspective. 

If  $(M,\omega)$ is a symplectic manifold, the Hamiltonian vector field  $X_F$ defined by
 the function $F$
 on $M$ is the solution of the equation
$$  i(X_F)\,\omega = dF
$$
A dynamics is given by choosing  a Hamiltonian $H$ through the corresponding Hamiltonian vector field $X_H$.

For instance, the phase space $M$ of the Hamiltonian systems
studied in classical mechanics is the cotangent bundle $T^*Q$ of
the configuration space $Q$ \cite{AbrMarsden}--\cite{Vilasi}.
They  are endowed with a canonical  exact symplectic structure $\omega_0$ that
in cotangent bundle  coordinates $(q_i,p_i)$,  $i=1,\ldots,n$, is given by
$\omega_0 =  dq_i\wedge dp_i $, i.e. $\omega_0 = - d\theta_0$,   with $\theta_0 = p_i\,dq_i$,
where the summation on repeated indices is understood. The Lagrangian formulation
 for regular Lagrangian functions $L$ is also an example. The manifold is the tangent bundle  $M=TQ$  
 and  the symplectic structure is not canonical  but depends on the choice of the function $L$. 

The Poisson bracket of two functions
$F_1$ and $F_2$ in a  symplectic manifold $(M,\omega)$  is defined as the symplectic product of the
corresponding Hamiltonian vector fields
$$
 \{F_1,F_2\} = \om_0(X_{F_1}\,,X_{F_2}) =X_{F_2}F_1\,.
$$
It can be verified that, when written in Darboux coordinates, the
integral curves of $X_H$ are solutions of Hamilton equation
and for the Poisson brackets we recover the expression usually
given in classical mechanics.

 The flow of the
Hamiltonian vector field $X_H$ on a symplectic manifold
$(M,\omega)$
commutes with the action of 
$X_H$;  thus,   if $F$ is a function in $M$ we have
$$
 \phi_t^*(X_HF) = X_H (\phi_t^*F) =  \frac d{ds}[\phi_s^* ( \phi_t^*F)]_{|s=0}
 = \frac d{ds}(\phi_{s+t}^*F)_{|s=0} = \frac d{du}[\phi_u^*F]_{|u=t}  \,.
$$
For a given function $G$ which is going
to play the r\^ ole of the  function (\ref{clausius}) we obtain
$$
 \frac{d}{dt}(\phi_t^*G)=\phi_t^*(\{G,H\})=-\phi_t^*(X_GH) \,,
$$
and integrating this relation with respect to  the time  from
$t=0$ to $t=T$, we arrive at
\begin{equation}
 \frac{1}{T}[G\circ \phi_T-G]=-\frac{1}{T}\int_0^T(X_GH)\circ\phi_t\, dt =
 \frac{1}{T}\int_0^T\{G,H\}\circ\phi_t\, dt.\label{virrel}
\end{equation}

 This result is sometimes known as the hypervirial theorem \cite{Hi60}:
If either the motion is periodic of period $T$, or when
$G$  remains bounded in its time evolution, taking the limit
$T\to\infty $:
\begin{equation}
\d<\{G,H\}> = 0 \,.  \label{promedioPP}
\end{equation}
Actually a family of hypervirial theorems was introduced in
\cite{Hi60} with an especial emphasis on homogeneous first-degree
functions in momenta, but the formulation is still valid for any
choice of the function $G$.

 In the particular case $Q=\mathbb{R}^3$
the phase space is the cotangent bundle $M=T^*\mathbb{R}^3$ which
is endowed with its canonical
symplectic structure $\omega_0$.  Then the Hamiltonian vector
field $X_G$ of the Clausius function $G$ given by  $G({\bf r},
{\bf p}) = {\bf r}\cdot {\bf p}$, takes the form
$$
 X_G = \sum_{i=1}^3\left( x^i\,\pd{}{x^i} - p_i\,\pd{}{p_i}\right),
$$
and represents the generator of dilations in $M=T^*\mathbb{R}^3$:
the infinitesimal dilations in $Q = \mathbb{R}^3$ are generated
by  $D = {\displaystyle \sum_{i=1}^3 }x^i\,\partial/\partial
{x^i}$ and its cotangent  lift to the phase space  as an infinitesimal generator of point
transformations, gives us $X_G$.

When   $H$ is a natural Hamiltonian (kinetic term plus a potential)
$$
 H({\bf r}, {\bf p}) = \frac{1}{2m} \,{\bf p}^2 + V({\bf r})
 = H_0({\bf p}) + V({\bf r}),
$$
the action of $X_G$ on $H$ becomes
$
 (X_GH)({\bf r}, {\bf p}) = - 2 H_0({\bf p}) + {\bf r}\cdot \boldsymbol{\nabla}V.
$
Therefore, the relation (\ref{virrel}) is 
$$
 \frac{1}{T}[G\circ \phi_T-G] = \d< 2 H_0>-\d< {\bf r}\cdot \boldsymbol{\nabla}V>,
$$
and, in a periodic motion of period $T$, or in the limit $T\to\infty $ when  $G$ remains bounded
in its time evolution, (\ref{promedioPP}) 
reduces to 
 the standard result in the Hamiltonian framework
$$
 \d< 2 H_0> = \d< {\bf r}\cdot \boldsymbol{\nabla}V>.
$$

Summarizing,  the standard virial and hypervirial theorems can be
considered as particular cases of more general properties stated
for Hamiltonian systems in symplectic manifolds.

\section{The virial theorem in the Lagrangian formalism}

It has been shown  (see \cite{Cr81,Cr83}) that a (regular) Lagrangian system is a
particular case of a Hamiltonian system on a symplectic manifold,
on the tangent bundle $TQ$ of the
configuration space $Q$ and with a symplectic structure depending on  
 the Lagrangian function. We can therefore translate the theory developed in the
 Hamltonian formalism to a Lagrangian framework.
 
  Two important geometric ingredients are
the Liouville vector field $\Delta$, that is the generator of
dilations along the fibres, and the vertical endomorphism $S$.
Given a differentiable
function $L$ on $TQ$, we can construct a semibasic 1-form
$\theta_L\in\bigwedge^1(TQ)$, an exact two-form
$\omega_L\in\bigwedge^2(TQ)$ and an Energy function by
$$
  \theta_L = S^*(dL)\,,{\quad} \omega_L = - d\theta_L  \,,{\quad}
  E_L=\Delta(L)-L \,.
$$
If the Lagrangian $L$ is regular then $\omega_L$ is symplectic
and the Lagrangian dynamics is given by the uniquely
determined vector field
$X_L$  solution of the equation
$$
 i(X_L)\omega_L = dE_L \,.
$$
$X_L$  satisfies the second-order property $S(X_L)=\Delta$,  and the
curves on $Q$ that are a  projection of the integrals curves of $X_L$  in $TQ$
satisfy the second-order  Euler--Lagrange equations.

To summarize,  a regular Lagrangian $L$
determines a symplectic structure $\omega_L$  in $TQ$ and 
the Lagrangian formalism is an instance of the theory
of  Hamiltonian dynamical systems.
The virial theorem reduces in this case to
\begin{equation}
 \frac{1}{T}[G\circ \phi_T-G] = - \frac{1}{T}\int_0^T(X_G(E_L))\circ\phi_t\, dt
 = \frac{1}{T}\int_0^T\{G,E_L\}_L\circ\phi_t\, dt,\label{relVTL}
\end{equation}
where $X_G$ is the vector field such that $i(X_G)\,\omega_L = dG$,
and $\{G,E_L\}_L = \omega_L(X_G,X_{L})$.  If the motion is
periodic of period $T$, or if the function $G$  remains bounded in
its time evolution, when taking  the limit of  $T$ going
to infinity we obtain
\begin{equation}
  \d<X_G(E_L)>=0.
\end{equation}

Let us consider  the simple case of a natural Lagrangian defined
in $Q=\mathbb{R}^3$
$$
  L({\bf r}, {\bf v}) =  {\smallonehalf} \,m\,{\bf v}^2 -  V({\bf r}) \,.
$$
The energy function and the symplectic form are given by
$$
  E_L({\bf r},{\bf v}) = {\smallonehalf} \, m\,{\bf v}^2 + V({\bf r}),\qquad  \omega _L
  = m\,d{\bf r}\wedge d{\bf v},
$$
and when $G$ is the observable function $G({\bf r},{\bf v})=m\, {\bf r}\cdot {\bf v}$, then
$$
 X_G = \sum_{i=1}^3\left( x^i\,\pd{}{x^i} - v^i\,\pd{}{v^i}\right),
 $$
or written in a simpler way
$$
X_G({\bf r}, {\bf v})={\bf r}\cdot {\boldsymbol{\nabla}}_{\bf r} -  {\bf v}\cdot {\boldsymbol{\nabla}}_{\bf v}.
$$
Note that here the minus sign shows that $X_G$ is the Hamiltonian
vector field defined by $G$ but it is not the lift to the tangent
bundle $TQ$ of the infinitesimal  generator of dilations in
$Q=\mathbb{R}^3$, but $X_G$  depends on the function $L$. Then we have
$$
(X_G(E_L))({\bf r},{\bf v})={\bf r}\cdot {\boldsymbol{\nabla}}V({\bf r})-{\bf v}\cdot (m\,{\bf v}) \,,
$$
and we thus recover the original virial theorem (\ref{sVTcf}).

One-parameter groups of point symmetries of the Lagrangian lead to
constants of motion but  
infinitesimal transformations that are not symmetries of $L$ can
also play a r\^ole in establishing virial theorems. Next we give a
geometric approach to  a result of \cite{vKa72}, whivh  is a
particular case of a more general result to be given in next
section where one-parameter groups of non-strictly canonical
transformations are to  be considered.

It is important to remember that that if $\phi$ is a diffeomorphism in $TQ$ that is obtained
from a diffeomorphism  $\varphi$ in the base $Q$,  that is
$\phi=T\varphi$, then the following two properties
\cite{Cr83} are satisfied
$$
 \phi^*\theta_L=\theta_{\phi^*L} \,, {\quad}  \phi^*E_L=E_{\phi^*L} \,.
$$
 Correspondingly, at the infinitesimal level, for a vector field $X\in \mathfrak{X}(TQ)$ that is a complete lift, $X=Y^c$, of a vector field in the base,  $Y\in  \mathfrak{X}(Q)$, we have
$$
 \mathcal{L}_X\theta_L=\theta_{X(L)} \,, {\quad} X(E_L) = E_{X(L)},
$$
where $\mathcal{L}_X$ denotes de Lie derivative with respect to $X$.

 In the following, given a 1-form $\alpha\in\bigwedge^1(Q)$ we denote by  $\widehat{\alpha}$ a  function on $TQ$
that is linear in the fibres defined by
$\widehat{\alpha}(q,v) = \<\alpha_q,v>$.
 Then, as $\theta_{\widehat \alpha}$ is the pull-back of $\alpha$ and $E_{{\widehat \alpha}}=0$,  $
 L' = L + \widehat{\alpha} $
defines the same Hamiltonian system on $TQ$ as $L$ if (and only if)  $\alpha $ is closed.
Hence a
vector field  $X=Y^c$, $Y\in  \mathfrak{X}(Q)$,  such that
$
  X(L) = \widehat \alpha$, 
with $\alpha$  a closed form  in $Q$,   is a symmetry of the
Hamiltonian dynamical system $(TQ,\omega_L,E_L)$ defined by $L$.
We can at  least locally write $\alpha=d\,\widetilde{h}$ where
$\widetilde{h}$ is the pullback through the tangent bundle
projection of a function $h$  in $Q$. Note that, making use of
this notation,   the property
$\widehat{dh}=\Gamma (\widetilde{h})$, is true for any vector field
$\Gamma$ on $TQ$ satisfying the second order differential equation                                                                                condition, $S(\Gamma)=\Delta$.

\begin{teorema}  Let $L$ be a regular Lagrangian and $X=Y^c$ a vector field
on $TQ$ such that $X(L) = a\,L+\widehat{dh}$, with
$a\in\mathbb{R}$. Then:
\begin{itemize}
 \item[(i)] the vector field  $X$ is a symmetry of the dynamical vector field $X_L$;
 \item[(ii)] the function $G=i(X)\theta_L-\widetilde h$ is such that  $X_L(G)=a\, L$.
\end{itemize}
\end{teorema}

\noindent{\sl Proof.} {\it  (i)} First, we have
$  \theta_{X(L)}=a\,\theta_L+d\,\widetilde {h}$, 
and therefore  $\omega_{X(L)}=a\,\omega_L$. Furthermore, the energy
$E_{X(L)}$ is given by  $E_{X(L)}=a\, E_L$. Consequently $
\mathcal{L}_X\omega_L=\omega_{X(L)}=a\,\omega_L$ and  $
\mathcal{L}_XE_L=E_{X(L)}=a\,E_L$.  Then we have
$$
  i([X,X_L])\omega_L  =  \mathcal{L}_X \bigl( i(X_L)\omega_L\bigr)
  - i(X_L) \bigl(\mathcal{L}_X\omega_L\bigr)
  =  \mathcal{L}_X \bigl( dE_L \bigr)   - i(X_L) \bigl(  a\,\omega_L\bigr) = 0\ .
$$
Hence $[X,X_L]$ is in the kernel of  $\omega_L$ and, as $\omega_L$
is symplectic (and therefore  regular), the kernel is trivial and
we arrive  at $[X,X_L]=0$.

{\it  (ii)} Since $X$ commutes with $X_L$ we then have
$\mathcal{L}_{X_L}(i(X)\theta_L) = i(X)(\mathcal{L}_{X_L}
\theta_L)$. Therefore
$$
\mathcal{L}_{X_L}\bigl(i(X)\theta_L\bigr) = i(X)(\mathcal{L}_{X_L}\theta_L)
 = i(X)dL=X(L)=a\,L+\widehat{dh} = a\,L+ X_L(h),
$$
and from here we get
$$
 X_L(G) = X_L \bigl(i(X)\theta_L-h\bigr) = a\,L \,.
$$

The result  of this theorem can be used to obtain the
following virial type relation
$$
 \frac1{t_2-t_1}\bigl[G(t_2)-G(t_1)\bigr] = \frac a{t_2-t_1}\int_{t_1}^{t_2} X_L(G)\, dt
 =  \frac a{t_2-t_1}\int_{t_1}^{t_2} L\, dt \,,
$$
and consequently, in a periodic motion of period $T$ we can take
$t_1=0$ and $t_2=T$  an we obtain $\d<L>=0$. In the more general
case,
$$\frac1{t_2-t_1}[G(t_2)-G(t_1)] = a\, \d<L>,
$$
where $L$ is averaged in $t\in [t_1, t_2]$ and we can also take
the limit when   $t_1\to-\infty$, $t_2\to\infty$.

  As an example  we consider the vector field $X$ that is the
complete lift of the vector field $Y$ generating dilations in $Q$,
given by
$$
   X = Y^c= \sum_{i=1}^3\left( x^i\,\pd{}{x^i} + v^i\,\pd{}{v^i}\right) \,.
$$
Then the action of $X$ on the Lagrangian $L$ of the harmonic
oscillator is given by
$$
 X(L) =  X\left({\smallonehalf} \,m\,{\bf v}^2 - {\smallonehalf} \,m\,\omega^2\, {\bf r}^2\right)   = 2\,L,
$$
so that $X$ is a symmetry of dynamical vector field of the
harmonic oscillator with $a=2$.  Hence we have the following
property
\begin{equation}
 \frac{1}{t_2-t_1}\left[\,\sum_{i=1}^3x_i\pd{L}{v^i}\,\right]_{t_1}^{t_2} = 2\, \d<L>,\label{sVFL}
\end{equation}
where on the right hand side $L$ is averaged in $t\in [t_1, t_2]$.
Once again we can particularize the time interval $t_2-t_1$ for a
periodic $T$ of the system  (this result is also true for the
limit $t_2$ going to infinity since the motion is always bounded
and periodic).

 This example can be obtained  as a particular case when starting with
the general one-dimensional case of a natural Lagrangian $L$ given
by
$$   L(x,v)  = {\smallonehalf} \,m\, v^2 - V(x) \,.
$$
Let   $Y$ be the  vector field in $\mathbb{R}$ given by
\begin{equation}
  Y = \xi(x)\pd{}x       \label{vfY}
\end{equation}
whose  complete lift $X$ is given by
\begin{equation}
  X(x,v) = Y^c(x,v) = \xi(x)\pd{}x + \Bigl(v\pd{\xi}x\Bigr)\pd{}v,  \label{vfYc}
\end{equation}
then we have
$$  (X(L))(x,v)=-\xi(x) \,V'(x) +m\,v\,\xi'(x) \,v,
$$
and therefore the condition $Y^cL=a\,L$, with $a\in \mathbb{R}$ is
written
$$
 \left\{\begin{array}{rl} &2\xi'(x) =a,\\
 &\xi(x)\, V'(x) = a\,V(x)\end{array}  \right.
$$
 from where we obtain
$$  \xi(x)=\frac 12 a\, x+C, \qquad 
V(x)=C_2(x+C_1)^2,
$$
and we recover as a particular case for $a=2$ the above mentioned
harmonic oscillator.

\section{The virial theorem for position dependent mass systems}
\indent

The study of systems with position-dependent mass is receiving a
lot of attention recently, with  the non-commutativity of
mass and momentum being a difficulty to be taken into account  in
the quantization process  \cite{vR,Le95,KK03}. From the classical
point of view the situation corresponds to systems for which the
kinetic energy is defined from a non-Euclidean metric. Therefore
the systems are not invariant under dilation and one should look
for alternative functions giving  rise to a virial-like theorem.
The more general theory we have developed here allows us to deal with
 this more general case.

We now consider the one-dimensional case of a Lagrangian $L$ given
by
\begin{equation}
  L(x,v) =   {\smallonehalf} \, m(x) v^2 - V(x),\label{Lmdep}
\end{equation}
where $m(x)$ is a positive differentiable function.

According to the formalism presented in the preceding section,
we must look for a vector field $Y$ in $\mathbb{R}$ given by
(\ref{vfY}) and its corresponding complete lift $X=Y^c$
(\ref{vfYc}) such that $X(L)=a\,L$, with $a\in \mathbb{R}$. Then as we have
$$
  (X(L))(x,v)=\frac 12\, \xi(x) \,m'(x) v^2-\xi(x) \,V'(x) +v\,\xi'(x) \,m(x)\,v,
$$
the mentioned condition leads to
$$   \left\{\begin{array}{rl}
 &2\xi'(x) +\xi(x)\, \mu(x) = a,\\
 &\xi(x)\, V'(x)=a\,V(x) \end{array}   \right.
$$
where $\mu(x)=m'(x)/m(x)$.

 The first equation is an inhomogeneous linear differential equation with a general
solution 
\begin{equation}
 \xi(x)=\frac 1{\sqrt{m(x)}}\left(C_1+\frac a2\int_0^x\sqrt{m(\zeta)}\, d\zeta\right).\label{xi}
\end{equation}
Consequently,  the vector field $Y$  is a sum of two vector
fields. The first one $Y_1$, which is obtained putting $a=0$ in
the preceding expression, preserves the kinetic term $T$ and it therefore 
represents a  Killing vector of the associated metric.
The second one $Y_2$ is given by
$$
  Y_2=\frac a2 \frac 1{\sqrt{m(x)}}\,\left(\int_0^x\sqrt{m(\zeta)}\, d\zeta\right)\pd{}x\,.
$$
The potential energy $V(x)$ is then given by
\begin{equation}
  V(x)=C_2\exp\left(\int^x \frac a{\xi(\zeta)}\, d\zeta\right) ,\label{vaL}
\end{equation}
while the function $G$ is 
\begin{equation}
  G(x,v) = m(x)\,\xi(x)\,v  = \sqrt{m(x)} \left(C_1+\frac a2\int_0^x\sqrt{m(\zeta)}\, d\zeta\right) v \,.      \label{Gmd}
\end{equation}
and the virial theorem provides the relation
$$  \d<X_L(G)>=a\,\d<L>.
$$

As an example we can consider  the differential equation for a 1-dimensional
nonlinear oscillator studied in 1974 by Mathews and Lakshmanan
\cite{ML}
$$
  (1 +\lambda q^2)\,\ddot{q}-\lambda\,q\,\dot{q}^2 + \alpha^2\,q  = 0
  \,,\quad\lambda>0\,.
$$
The  general solution takes the form $q(t) = A \sin(\omega\,t + \phi) \,,$ with the following additional restriction linking frequency and amplitude
$$
  \omega^2  = \frac{\alpha^2}{1 + \lambda\,A^2} \,.
$$
The system admits a Lagrangian formulation with Lagrangian:
$$
  L_\lambda(q,\dot q)  = \frac{1}{2}\ \frac{1}{1 + \lambda\,q^2} \,(\dot{q}^2 -
\alpha^2\,q^2)\ .
$$
It is a system with nonlinear oscillations with a frequency (or
period) showing amplitude dependence. We can also allow negative
values for $\lambda$,  but when $\lambda<0$  the values of $x$ are
limited by the condition $|x|<1/\sqrt{|\lambda|}$ \cite{CRS04}.

In the limit $\lambda\to 0$ we recover the equation of motion and
the Lagrangian of the harmonic oscillator and the frequency
becomes independent of the amplitude. In this sense the system can
be seen as a deformation of the harmonic oscillator. It  can also
be considered as an oscillator with a position-dependent effective
mass which depends on $\lambda$,  $m_\lambda(q)= (1 +
\lambda\,q^2)^{-1}$. A quantum version of this model was studied
in \cite{CRS04}, and a superintegrable  generalization to several
degrees of freedom in classical mechanics was proposed in
\cite{CRSS04} and the corresponding quantum version  was studied
in \cite{CRS07a}-\cite{CRS07c}.

In order to deal simultaneously with all possible values of
$\lambda$ it is convenient to  introduce the following
$\kappa$-trigonometric functions, where   $\kappa\in \mathbb{R}$:
$$
  \Cos_{\k}(x) = \left\{\begin{array}{ll}
    \cos{\sqrt{\kappa}\,x}       &{\rm if}\  \kappa>0, \cr
    {\quad}  1               &{\rm if}\  \kappa=0, \cr
    \cosh\!{\sqrt{-\kappa}\,x}   &{\rm if}\ \kappa<0, \end{array}
\right.{\qquad}
  \Sin_{\k}(x) = \left\{\begin{array}{ll}
    \frac{1}{\sqrt{\kappa}} \sin{\sqrt{\kappa}\,x}     &{\rm if}\ \kappa>0, \cr
    {\quad}   x                                &{\rm if}\ \kappa=0, \cr
    \frac{1}{\sqrt{-\kappa}}\sinh\!{\sqrt{-\kappa}\,x} &{\rm if}\   \kappa<0,
  \end{array}\right.
$$
and the $\kappa$-dependent tangent function $ {\rm T\,}_\kappa(x)$
defined in the natural way,  $ {\rm T\,}_\kappa(x) =  {\rm
S\,}_{\kappa}(x)/ {\rm C\,}_{\kappa}(x)$.  The fundamental
properties of these curvature-dependent trigonometric-hyperbolic
functions are
$$
  \Cos_{\k}^2(x) + \kappa \, \Sin_{\k}^2(x) =1 \,,
$$
and
$$
  \Cos_{\k}(2x) = \Cos_{\k}^2(x) - \kappa \, \Sin_{\k}^2(x)\,, {\qquad}
   \Sin_{\k}(2x) = 2\Sin_{\k}(x) \Cos_{\k}(x)\,.
 $$
It can also verified that the derivatives of these functions are
given by
$$
  \frac{d}{dx}  \Sin_{\k}(x) =  \Cos_{\k}(x) \,, {\qquad}
  \frac{d}{dx}  \Cos_{\k}(x) = - \kappa \Sin_{\k}(x)  \,,
$$
as well as
$$
\frac{d}{dx}{\rm T\,}_{\kappa}(x)= \frac 1{{\rm C\,}_{\kappa}^2(x)}\,, {\qquad}\frac{d}{dx}{\rm T\,}_{\kappa}^{-1}(x)=\frac 1{1+\kappa\, x^2}.
$$

In the  case we are considering, the position dependent mass
$m(q)$ is given by $m(q)=({1 + \lambda\,{q}^2})^{-1}$ and the
function $\xi(q)$ becomes
$$
  \xi(q)=\sqrt{1 + \lambda\,q^2}\left(C_1+\frac a2\int^q\frac {dq'}{\sqrt{1 + \lambda\,{q'}^2}}\, \right) \,.
$$
Under the change of variables $q = \Sin_{\k}(u)$ with
$\kappa=-\lambda$, and therefore  $\sqrt{1 + \lambda\,q^2} =
\Cos_{\k}(u)$,  we obtain
$$
  \xi(u) = \Cos_{\k}(u)\left( C_1+\frac a2 \,u\right),
$$
or rewritten in terms of the original variables
$$
 \xi(q)=\sqrt{1 + \lambda\,q^2}\,\left( C_1+\frac a2 \,\Sin_{\k}^{-1}(q)\right).
$$

 The potential energy $V(q)$ is then given by the integral
$$
  V(q )= C\exp\left(\int^q \frac a{\xi(\zeta)}\, d\zeta\right)
   = C\exp\left(\int_0^u\frac a{C_1+\frac a2 \,\zeta}\,d\zeta\right),
$$
that leads to the value
$$
  V(q)  = C\left(1+\frac a{2C_1}\, u\right)^2
  = C\left(1+\frac a{2C_1}\, \Sin_{\k}^{-1}(q)\right)^2.
$$
For such a potential the function $G$ is given by
$$
G(q,v) = \frac v{\sqrt{1 + \lambda\,q^2}}\left( C_1+\frac a2 \,\Sin_{\k}^{-1}(q)\right),
$$
and
the virial theorem provides the relation
$$  \d<X_L(G)>=a\,\d<L>.
$$

 We close this section with a comment on the general case of the Lagrangian
(\ref{Lmdep}).   If we introduce a new coordinate $u$ given by the
relation
$$
  u(x) =\int_0^x\sqrt{m(\zeta)}\, d\zeta
$$
and such that $du/dx=\sqrt{m(x)}$, then the expression for the
function $\xi(x)$ becomes
$$
 \xi = \frac 1{\sqrt{m(x)}}\left(C_1+\frac 12 u(x)\right) \,,
$$
and the potential $V(x)$ is just a quadratic function in the function $u(x)$
$$
 V(x)=C_2\left(1+\frac a{2C_1} \,u(x)\right)^2   \,.
$$

\section{The virial theorem and non-strictly canonical transformations}

In classical mechanics a transformation that  preserves the
Poisson brackets up to a multiplicative constant,  called the
valence of the transformation, is said to be a 
canonical transformation \cite{SaletanCromer}  (if the valence is
the unity the transformation is strictly canonical, otherwise is non-strictly canonical). In
differential geometric terms, a vector field $X$ on a symplectic
manifold $(M,\omega )$  is the generator of a one-parameter group
of non-strictly canonical transformations if there exists a real
number $a\ne 0$ such that $\mathcal{L}_{X}\omega=a\,\omega$.

Given such a vector field $X$, let us choose a vector field $X_1$  such that
  \begin{equation}
  \mathcal{L}_{X_1}\omega=-\omega.\label{X1}
\end{equation}
For instance, when  $(M,\omega=-d\theta)$ is  an exact symplectic
manifold the vector field  $X_1$ can be chosen to be defined by
(see \cite{NaVV86})
$$
  i(X_1)\omega=\theta,
$$
because then
$$
  \mathcal{L}_{X_1}\theta = i(X_1)d\theta + d(i(X_1)\theta)
  = - i(X_1)\omega + d\bigl(i(X_1)\theta\bigr) = - \theta + d(i(X_1)\theta) ,
$$
and therefore
$$
  \mathcal{L}_{X_1}\omega=-\mathcal{L}_{X_1}(d\theta)=-d\mathcal{L}_{X_1}\theta=d\theta=-\omega.
$$
Note, however, that manifolds exist  that do not admit exact symplectic forms, because of topological obstructions. For instance, 
there is no exact volume form in a compact manifold  $M$, as a consequence of Stokes theorem, while if $\omega $ is an exact symplectic form in a $2n$-dimensional manifold $M$, then $\omega^{\wedge n}$ would be an exact volume form in $M$.
 
Anyway, if $X_1$ satisfies (\ref{X1}), then   $X+a\,X_1$  is a locally-Hamiltonian vector field since it satisfies
$$
  \mathcal{L}_{X+a\,X_1}\omega=a\,\omega -a\, \omega =0.
$$
That means the existence of a closed 1-form $\alpha$ such that
$$
  i(X)\omega+a\, i(X_1)\,\omega=\alpha.
$$
Conversely, given a closed 1-form $\alpha$ the preceding relation
defines a vector field generating  a one-parameter (local) group
of non-strictly canonical transformations $\phi_\epsilon$ with
valence $e^{a\epsilon}$.

In a Darboux chart for which
$$
  \omega = \sum_{i=1}^n dq^i\wedge dp_i  \,,
$$
we can choose as $\theta$ (such that $\omega=-d\theta$) the 1-form given by
$$
 \theta = \frac{1}{2} \sum_{i=1}^n \left( p_i\, dq^i-q^i\, dp_i\right),
$$
and then the vector field $X_1$ is the dilation generator
$$
X_1=-\frac 12\sum_{i=1}^n\left( q^i\, \pd{}{q^i}+p_i\, \pd{}{p_i}\right) \,.
$$
If $\alpha =d\phi$ (at least locally), then we can write $X$  as follows
$$
  X = X_\phi - a\,X_1  \,,
$$
and then the action of $X$ on the Hamiltonian is given by
$$
  X(H) = \{H,\phi\} - a\,X_1(H) \,.
$$
Using a Darboux chart,  if we assume that   $H(q,p)=H_0(p)+V(q)$,  with $H_0$ a quadratic function, we find
$$
  X_1(H) = - \frac{1}{2} \left(2\,H_0(p)+\sum_{i=1}^nq^i\pd V{q^i}\right)
$$
and taking into account that
$$
  \left\{H,\sum_{k=1}^nq^k\,p_k\right\} =  \sum_{k=1}^nq^k\pd V{q^k}-\sum_{k=1}^np_k\pd{H_0}{p_k}
   =  \left(\sum_{k=1}^nq^k\pd V{q^k}\right) - 2\,H_0(p)
$$
we obtain
$$
X(H) = \left\{H,\phi+\frac a2\sum_{i=1}^nq^ip_i\right\}+2\, a\, H_0 \,.
$$
Now integrating in time from $t=0$ to $t=T$ and, as according to
(\ref{promedioPP}) the average of the Poisson bracket vanishes,
we obtain, in the limit of $T$ going to infinity (or when the
motion is periodic with periodic $T$), the following equality
$$
  2\, a\, \d<H_0>=\d<X(H)> \,.
$$
Finally,   if there exists $b\in \mathbb{R}$ such that $X(H)=b\,
H$, then $\{H,\phi\}-a\,X_1(H) = b\, H$, and we arrive at
$$
 2\, a\, \d<H_0>=b\,E .
$$

As a concrete example, let us consider a homogeneous potential of
degree $d\ne 2$, i.e.
$$\sum_{k=1}^nq^k \pd V{q^k} = d\, V,\qquad d\ne 2,
$$
and the one-parameter group of transformations generated by
$$
  X_a = \sum_{i=1}^n\left( \frac{a-1}2 q^i\, \pd{}{q^i}+ \frac{a+1}2p_i\, \pd{}{p_i}\right) \,.
$$
One immediately computes
$$    \mathcal{L}_{X_a}\omega = a\,\omega
$$
and, therefore, $X_a$ generates non-strictly canonical
transformations. Besides, if we chose $a=(d+2)/(d -2)$ we have
$$
  X_a(H) = \frac{2\,d}{d-2}H
$$
i.e. $b=2d/(d-2)$, and we are in the situation described before.  Hence, using the
previous result we have
$$
  (d+2)\langle\langle H_0 \rangle\rangle= d\, E,
$$
or equivalently
$$
  2\langle\langle H_0 \rangle\rangle = d \langle\langle V \rangle\rangle,
$$
which is the standard form of the virial theorem for homogeneous
potentials. Another example combining scaling of momenta with translations in coordinates   was used for the Toda lattice in \cite{NaVV86}. 

Finally, it is to be remarked that the case we considered in the
preceding section of infinitesimal groups  of point
transformations such that $X(L) = a\, L$ (up to a gauge  term)  is
only a particular case (with $a=b$) of the preceding situation  for the
Hamiltonian system $(TQ,\omega_L,E_L)$.

\section{The virial theorem in quantum mechanics}
\indent

There has been an increasing interest in 
 the geometric formulation of quantum mechanics
 as a particular case of a 
Hamiltonian dynamical system (see e.g.
\cite{BoCG91}-\cite{CaClMa07b} and the references therein), the main
difference with the classical mechanics case being  that the
symplectic manifold is a (may be infinite-dimensional) Hilbert
space seen as a real Banach space. More specifically,  a separable
complex Hilbert space  $(\mathcal{H} ,\<\cdot,\cdot >)$  can be
considered as a  real linear space, 
 then denoted
$\mathcal{H}_{\mathbb{R}}$. The  norm in $\Hil$ defines a norm in
$\mathcal{H}_{\mathbb{R}}$, where
$\|v\|_{{\mathbb{R}}}=\|v\|_{\mathbb{C}}$. Of course, more accurately one knows 
that  pure states in quantum mechanics are rays rather than vectors, but this fact can also be taken into account by replacing the Hilbert space for the corresponding projective Hilbert space.

The  real  linear space $\mathcal{H}_{{\mathbb{R}}}$ is  endowed
with a natural symplectic structure as follows:
$$
  \omega (u,v) = 2 \,{\rm Im\,} \<u,v>.
$$
In fact, $\omega$ is a skew-symmetric real bilinear map and the ${{\mathbb{R}}}$-linear map  $\widehat\omega: \mathcal{H}_{{\mathbb{R}}}\to \mathcal{H}^*_{{\mathbb{R}}}$ defined by   $\widehat\omega(u)v = \omega(u,v)$   is not only  injective but it is also an isomorphism, because 
 if  $\widehat{\omega}(u)=0$, then  $\widehat\omega(u)(iu) = 2\,\<u,u>=0$,  and consequently  $u=0$.
Riesz theorem can be used to prove that the map is also surjective.

The  Hilbert $\Hil_{{\mathbb{R}}}$ can be  considered as a  real
manifold modelled by a Banach space admitting a global chart.
Moreover,  for each $v\in \Hil_{{\mathbb{R}}}$ the tangent space
$T_v\Hil_{{\mathbb{R}}}$ is canonically isomorphic to the own
$\Hil_{{\mathbb{R}}}$: we  associate with  $w\in
\Hil_{{\mathbb{R}}}$ the vector in the  tangent space
$T_v\Hil_{{\mathbb{R}}}$ defined by
$$
  \chi_v(w)f =\frac  d{dt}f(v+t\,w)_{|t=0},
$$
where the function $f$ is differentiable in a neighbourhood of the vector $v$, $f\in C^\infty(v)$.  This is an  isomorphism $\chi_v:\Hil_{{\mathbb{R}}}\to T_v\Hil_{{\mathbb{R}}}$  providing us an  identification  which corresponds to the one given by the free transitive  action of the Abelian group of  translations on $\Hil_{{\mathbb{R}}}$.

One can prove (see later on) that the constant symplectic
structure $\omega$ in $\Hil_{{\mathbb{R}}}$,  considered as a
Banach manifold,   is exact, i.e., there exists a 1-form
$\theta\in \bigwedge^1(\Hil_{{\mathbb{R}}})$ such that
$\omega=-d\theta$. Such a   1-form $\theta \in \bigwedge
^1(\mathcal{H})$  is, for instance, the one   defined by
$$
 \theta(v) [\chi_v(w)]=-{\rm Im\,}\<v,w>, \label{defomega}
$$
because then  $\omega=-d\theta$ is a symplectic 2-form such that
$$ \omega(v)(\chi_v(u),\chi_v(w))=2\,{\rm Im\,} \<u,w >.
$$

A  {\sl continuous\/} vector field in $\Hil_{{\mathbb{R}}}$  is a
{\sl con\-ti\-nuous}\, map $X\colon \mathcal{H}_{{\mathbb{R}}}\to
\mathcal{H}_{{\mathbb{R}}}$. For instance, for each  $v\in \Hil$,
the  constant vector field $X_v$ defined by
$$  X_v (w)=\chi_w(v),
$$
is the generator of the one-parameter subgroup of transformations
of $\Hil_{{\mathbb{R}}}$ given by
$$  \Phi(t,w)=w+t\, v\,,
$$
i.e. with the natural identification of  $T\Hil_{{\mathbb{R}}}$
with $\Hil_{\mathbb{R}}\times \Hil_{{\mathbb{R}}}$,
$$  X_v:w\mapsto (w,v)\,.
$$
The values at a point of such vector fields generate the tangent
space at the point.

Similarly, for each  vector $v\in \Hil_{{\mathbb{R}}}$ there is a
constant 1-form $\alpha_v$ in $\Hil_{{\mathbb{R}}}$, given by
$$  \alpha_v: w\mapsto\<v,w>\,.
$$
Obviously,
$$  \alpha_{v_1}(X_{v_2})=\<v_1,v_2>\,,
$$
and therefore
$$
 \alpha_{v_1+\lambda\, v_2}= \alpha_{v_1}+\lambda \alpha_{v_2} \,,{\quad}
 \forall \lambda\in {{\mathbb{R}}}\,.
$$
 The   1-form $\theta$  defined above  satisfies
$$  \theta(X_v)=-{\rm Im\,}\<\cdot,v>\,,
$$
because according to the definition of the 1-form $\theta$,
$$ [\theta(X_v)](w) = \theta(w)[X_v(w)]
 = \theta(w)[\chi_w(v)] = -{\rm Im\,}\<w,v>\,.
$$
One can see that  $X_w[\theta(X_v)]$ takes a constant value:
$$
 X_w[\theta(X_v)](u) = \frac{d}{dt}\left[\theta(X_v)(u+tw)\right]_{|{t=0}}
 = - {\rm Im\,}\frac  d{dt}\<u+tw,v>_{|{t=0}} = -{\rm Im\,}\<w,v>\,.
$$
This allows us to check that $\omega=-d\theta$, because for any
pair  $v,w\in \Hil$, as $X_v$ and  $X_w$ commute, $[X_v,X_w]=0$,
we have
$$
  -d\theta(X_v,X_w) = -X_v\,\theta(X_w)+X_w\theta(X_v)
  =-2\,{\rm Im\,}\<w,v> = \omega(X_v,X_w)\,.
$$
As another particular example of vector field,  consider the vector
field  $X_A$ defined by the $\mathbb{C}$-linear map
$A:\mathcal{H}\to \mathcal{H}$, and in particular when  $A$ is
self-adjoint. With the natural identification of
$T\Hil_{{\mathbb{R}}}\approx \Hil_{{\mathbb{R}}}\times
\Hil_{{\mathbb{R}}}$, $X_A$ is given by
$$
 X_A:v\mapsto (v,Av)\in \Hil\times \Hil\,.
$$
When $A=I$ the vector field $X_I$ is the  Liouville generator of
dilations along the fibres, usually denoted
$\Delta=X_I$ and given by
$\Delta(v)=(v,v)$.

Given a self-adjoint operator  $A$ in $\Hil$ we can define a real
function in $\Hil_{{\mathbb{R}}}$ by
$$  a(v) =\<v, Av>\,,
$$
i.e. the evaluation map, which can be rewritten
$$  a=\<\Delta,X_A>\,.\label{defa}
$$
Then,
$$  \aligned
da_v(w )&=\frac d {dt} a(v+tw)_{|{t=0}}=\frac d {dt} \left[
\<v+tw,A(v+tw)>\right]_{|{t=0}}\\
&=2\, {\rm Re\,}\<w,Av>=2\, {\rm Im\,} \<-iAv,w>=\omega(-iAv,w).
\endaligned
$$
If we recall that the Hamiltonian vector field defined by the
function $a$ is such that for each $w\in T_v\Hil=\Hil$,
$$  da_v(w)=\omega(X_a(v),w)\,,
$$
we see that
$$
 X_a(v)=-i\,Av \,.
$$
Therefore if  $A$ is the  Hamiltonian  $H$ of a quantum system,
the Schr\"odinger equation describing time-evolution plays the
role of \lq Hamilton equations' for the Hamiltonian dynamical
system  $(\mathcal{H},\omega,h)$, where $h(v)=\<v, Hv>$: the
integral curves of   $X_h$ satisfy
$$
  \dot v=X_h(v)=-i\, Hv\,.
$$

The real functions $a(v)=\<v,Av>$ and $b(v)=\<v,Av>$ corresponding
to two selfadjoint operators $A$ and $B$ satisfy
$$
 \{a,b\}(v)=-i\,\<v,[A,B]v>\,,\label{Diraccor}
$$
because
$$
 \{a,b\}(v)=[\omega(X_a,X_b)](v)=\omega_v(X_a(v),X_b(v))=2\,{\rm Im\,}\<Av,Bv>\,,
$$
and taking into account that
$$
 2\,{\rm Im\,}\<Av,Bv> =- i\,[\<Av,Bv>-\<Bv,Av>]=-i\,[\<v,ABv>-\<v,BAv>]\,,
$$
we find the above result.

In particular, on the integral curves of the vector field $X_h$
defined by a Hamiltonian $H$,
$$  \dot a(v)=\{a,h\}(v)=-i\,\<v,[A,H]v>\,,
$$
that can be rewritten as
\begin{equation}
  \frac d{dt}\<v,Av> = -i\,\<v,[A,H]v>\,,  \label{EhTheorem}
\end{equation}
and is usually known as  the Ehrenfest theorem.
This is the starting point for the virial theorem in quantum mechanics.

Note that in the derivation of (\ref{EhTheorem}) we have used that
$Av$ belongs to the domain of $H$. This is not a restriction if
$H$ is defined in the whole Hilbert space, but in many  occasions
the self-adjoint Hamiltonian has a domain that is only dense in
$\mathcal H$ and if $A$ does not preserve this domain, extra
boundary terms should be added to the standard Ehrenfest theorem,
see refs. \cite{AbadEst91}-\cite{EstFalGir12} where the occurrence
of this kind of anomaly in quantum mechanics has been studied.
In this paper we shall assume that we have no boundary effects or,
in other words, the anomaly is absent and (\ref{EhTheorem}) holds.
Also note that identity (\ref{EhTheorem}) does not require that
$A$ be selfadjoint.

Finally, for the sake of completeness, we must comment that   if we want to consider the projective Hilbert space, one can start with 
the open submanifold ${\Hil}-\{ 0 \}$ of  $\Hil$  
and the action of the  two-dimensional Lie group $\mathbb{C}^*=\mathbb{C}-\{0\}$,
which is  an  Abelian Lie group isomorphic to the direct
product $\mathbb{R}_+\otimes U(1)$ of 
$\mathbb{R}_+=\{\lambda\in \mathbb{R}\mid \lambda>0\}$ and 
$U(1)=\{e^{i\varphi}\mid \varphi\in \mathbb{R}\}$. The corresponding vector fields are 
$X_d$ and  $X_{f}$ given by  
$$X_d (v)=\frac d{d\epsilon}
e^{-\epsilon }v_{ |\epsilon=0}=- v \,,
$$
$$
X_{f}(v)=\frac d{d\epsilon}
e^{-i\epsilon }v_{|\epsilon=0}=
-i\,  v 
\,.
$$
Let  $\Psi$  be  the map, 
\begin{equation}
\Psi:\Hil-\{0\}\to {\Bbb R}_+\times \SH\,,\qquad \Psi(v)=\left(\|v\|, 
\frac v{\|v\|}\right)\,.\label{difeoH}
\end{equation}
where $\SH$ is the subset of unit vectors of  $\Hil$
\begin{equation}
\SH=\{v\in \Hil\mid \<v,v>=1\} \,,\label{SHilbert}
\end{equation}
and denote  ${\rm pr}_1:\R_+\times \SH\to \R_+$ and  ${\rm pr}_2:\R_+\times
\SH\to  \SH$ the  projections of
$\Psi(\Hil-\{0\})=\R_+\times \SH$ on each factor. The map
${\rm pr}_2\circ\Psi:v\mapsto v/\|v\|$ identifies  $(\Hil-\{0\})/\mathbb{R}_+$ with $\SH$,
$(\Hil-\{0\})/{\mathbb{R}}_+\approx \SH$, and we obtain in this way a
reduced space which can be seen as a submanifold
$j_{\SH}:\SH\to \Hil$, endowed with the 2-form
$\omega_{\SH}=j_{\SH}^*\omega$, which is an exact ($ \omega_{\SH}=-d\theta_{\SH}$,  with  $\theta_{\SH}=j^*_{\SH}\theta$) but  degenerate 2-form.

As a second step in the reduction process as the symplectic action of the  Abelian group $U(1) = \{
e^{i\varphi}\mid \varphi\in {\mathbb{R}} \}$, on
  $\mathcal{H}$, 
preserves the submanifold  $\SH$, the set of orbits is
$\SH/U(1)$ is the projective Hilbert space ${\PH}=({\Hil}-\{0\})/\mathbb{C}^*$.
Note that the momentum map $J:\mathcal{H}\to\mathfrak{u}^*(1)$,  can be found to be given by 
$$\<J(v), i\lambda>=\lambda \<v,v>,$$
 i.e.  $J(v)=\<v,v>$ and therefore  $J^{-1}(1)= \SH$ is endowed with 
 a presymplectic form $\omega_{\SH}$ that is the pull-back of $\omega$. Its kernel is generated by the fundamental vector field corresponding to  $i\,1\in \mathfrak{u}(1)$.
 Consequently Marsden and   Weinstein reduction process leds to a uniquely determined reduced symplectic 
 form in the quotient space, the projective Hilbert space. Note however that such a
 symplectic form is not exact, recall that if dimension of $\Hil$ is finite its projective Hilbert space is compact.  Of course the evaluation map $a$ associated to the self-adjoint operator $A$ is not projectable and it must be replaced by the expectation value function \cite{CIMM} 
$$e_A(v)=\frac{\<v,Av>}{\<v,v>},
$$
which coincides with the evaluation map at points $v\in\SH$. The dynamical vector field obtained in this way is projectable and gives us the dynamics in the projective Hilbert space.

If the state $v$ is stationary then the above equation
(\ref{EhTheorem}) becomes an identity since both sides are zero.
In a generic state, if we integrate between $0$ and $T$ we obtain
$$  \<v(T),Av(T)>-\<v(0),Av(0)>=-i\int_0^T\<v(t),[A,H]v(t)>\,dt,
$$
and if $\<v(t),Av(t)>$ remains bounded, taking the limit when $T$
goes to infinity of the quotient of both sides by $T$ we find
$$  \d<{\<v,[A,H]v>}>=0,
$$
which is the quantum hypervirial theorem when no anomaly, due to
boundary terms, is present \cite{EstFalGir12}.

Suppose that  the Hamiltonian of a quantum system is
$$  H=\frac 12 {\bf P}\cdot {\bf P}+V({\bf X}),
$$
where ${\bf P}=-i\,\boldsymbol{\nabla}$ and natural units are used.
Let now $A$ be given by
$$
  A = \frac{i}{2} \left({\bf X}\cdot {\bf P}+{\bf P}\cdot {\bf X}\right).
$$
We first remark that $[{\bf X}\cdot {\bf P},H]
=[{\bf P}\cdot {\bf X},H]$ because of ${\bf X}\cdot {\bf P}-{\bf P}\cdot {\bf X}=3\,i $.

Taking into account the algebraic relation $[AB,C] =
A[B,C]+[A,C]B$ one can see that
$$
 [{\bf X}\cdot {\bf P}, \frac{1}{2} {\bf P}\cdot {\bf P}] =
 \frac{1}{2} \sum_{i=1}^3X_i\left[P_i\,,{\bf P}\cdot {\bf P}\right]
 +  \frac{1}{2} \sum_{i=1}^3\left[X_i\,, {\bf P}\cdot {\bf P}\right]\,P_i=i\,  \, {\bf P}\cdot {\bf P} \,,
$$
while
$$
[{\bf X}\cdot {\bf P},V({\bf X})]=\sum_{i=1}^3X_i\,[P_i,V({\bf X})] =
-i\,  \,{\bf X}\cdot{\boldsymbol{\nabla}}V({\bf X}),
$$
and therefore
$$
\<v,[{\bf X}\cdot {\bf P}, H]\, v>=i\,  \bigl(\<v,{\bf P}\cdot {\bf P}\, v>-\<v,({\bf X}\cdot{\boldsymbol{\nabla}}V({\bf X})) v>\bigr)  \,,
$$
and we obtain the standard quantum virial theorem
$$
\d<{\<v,{\bf P}\cdot {\bf P}\, v>}>-\d<{\<v,({\bf X}\cdot{\boldsymbol{\nabla}}V({\bf X})) v>} >= 0.
$$

Note that $A=\frac{i}2 \left({\bf X}\cdot {\bf P}+{\bf P}\cdot
{\bf X}\right)$ is the generator for the dilation subgroup.
Recall that if  a Lie group $G$ acts on $M$ on the left, then if
$\mu$ is a  $G$-invariant volume, we can define the so called
quasi-regular unitary representation in $(\mathcal{L}^2(M),\mu)$
as follows:
$$  (U(g)\psi)(x)=\psi(g^{-1}x).
$$
When $\mu$ is not $G$-invariant but quasi-invariant, in order to
get an unitary representation we must correct the right-hand side
by the square root of the Radon-Nikodym derivative.

In the case of dilations in the one-dimensional case,
$M={\mathbb{R}}$, and if $G$ is the dilation group there is no
invariant measure. The quasi-regular representation  then turns
out to be 
$$
[U(\lambda)\psi] (x)=\lambda^{-1/2}\psi(\lambda^{-1}x).
$$
The expression so defined is a one-parameter group of
transformations with canonical parameter $\alpha$ such that
$\lambda=e^\alpha$,
$$
 [U(\alpha)\psi] (x)=e^{-\alpha/2}\psi(e^{-\alpha}x),
$$
and its generator comes form
$$  \left.\pd{e^{-\alpha/2}\psi(e^{-\alpha}x)}{\alpha}\right|_{\alpha=0}=-\frac 12\psi (x)-x\, \pd{\psi}x,
$$
from where we deduce that the generator of this action is
$$  A = \frac 12+x\,\pd{}x=\frac{1}2\left(x\pd {}x+\pd{}xx\right).
$$

For $M={\mathbb{R}}^3$, the quasi-regular representation is
$$
[U(\lambda)\psi] ({\bf x})=\lambda^{-3/2}\psi(\lambda^{-1}{\bf x}),
$$
or in terms of the parameter $\alpha$,
$$\left.\pd{e^{-\alpha/2}\psi(e^{-\alpha}{\bf x})}{\alpha}\right|_{\alpha=0}=-\frac 32\psi ({\bf x})-{\bf x}\cdot {\boldsymbol{\nabla}}{\psi},
$$
and the generator of the action can be written
$$ A=\frac 32+{\bf x}\cdot {\boldsymbol{\nabla}}=\frac{1}2\left({\bf x}\cdot {\boldsymbol{\nabla}}+ {\boldsymbol{\nabla}}\cdot {\bf x}\right).
$$

In order to better understand the geometric properties of the virial 
theorem in quantum mechanics, it is interesting to study the quantum version
of the system with position dependent mass introduced in Sec. 4. 
Its quantization has been studied in detail in \cite{CRS04}, where
it was shown that the appropriate description of the system implies
the introduction of a non trivial metric $g=m(x){\rm d}x\otimes{\rm d}x$
in ${{\mathbb{R}}}$. Consequently,  the Hilbert space is $L^2(\mathbb{R},d\mu)$ with $d\mu = \sqrt{m(x)}\,dx$ and  then the norm of a function $\Psi$  is given by
\begin{equation}\label{norm}
||\Psi||^2 = \int_{-\infty}^\infty |\Psi(x)|^2 \sqrt{m(x)}\,{\rm d}x.
\end{equation}

The Hamiltonian operator of the system reads
$$  
H = H_0+V(x)=\frac12 P^2 +V(x) 
$$
where 
$$ P = -\frac{i}{\sqrt{m(x)}} \frac\partial{\partial x},
$$
is symmetric with respect to the scalar product 
induced by (\ref{norm}) and self-adjoint when appropriate boundary conditions are chosen.
Note also that $P$ is a Killing vector of the metric, i.e. 
$\mathcal{L}_P g=0$, and it  is, up to a factor $i$  the generator of the translation 
group in this non-homogeneous space. 

In order to apply the virial theorem we introduce the 
generator of the {\it dilation} group $A=\xi(x)\partial_x$ 
with the property 
$$[P,A]=\frac{a}2 P \,. 
$$
The general solution was obtained in (\ref{xi})
and is given by
\begin{equation}
 \xi(x)=\frac 1{\sqrt{m(x)}}\left(C_1+\frac a2\int_0^x\sqrt{m(\zeta)}\, d\zeta\right).
\end{equation}
Notice that the name of dilation group is justified because of the 
commutation relation of its generator with that of translations and 
also because of the action on the metric, i.e.
$${\cal L}_A g= a \,g,$$
that corresponds to a conformal transformation. 

We consider now the potential
 \begin{equation}
  V(x)=C_2\exp\left(-\int^x \frac b{\xi(\zeta)}\, d\zeta\right) ,
\end{equation}
that verifies
$$[V,A]=b\,V.$$
With all these ingredients we have
$$
\<v,[H, A]\, v>=a \<v,H_0\, v> + b \<v,  V\, v>,
$$
and consequently
$$
a \d<{\<v,H_0\, v>}>+b \d<{\<v,V v>} >= 0.
$$

Finally, some comments on Fock approach to the virial theorem in
quantum mechanics \cite{Fock30}:
                                           Starting with an arbitrary wave function $\phi$ we consider the
one-parameter family of trial functions
$\{\phi^\lambda=U(\lambda)\,\phi\mid \lambda\in \mathbb{R}_+\}$,
where $U$ is a linear representation of the dilation group.

 When $d=1$ the expectation value of the kinetic term $H_0$ is homogeneous of
degree $-2$ and the potential $V$ is assumed to be homogeneous of
degree $k$:
 $$\<\phi^\lambda, H_0\phi^\lambda>=\lambda^{-2}\, \<\phi, H_0\phi>,\qquad
 \<\phi^\lambda, V\phi^\lambda>=\lambda^{k}\, \<\phi, V\phi>,
$$
therefore
$$
 E_\lambda=\<\phi^\lambda, H\phi^\lambda>=\lambda^{-2}\, \<\phi, H_0\phi>+\lambda^{k}\, \<\phi, V\phi>.
$$

The best approach to an eigenvalue in the family will be by a
value of $\lambda$ such that
$$
  \frac {dE_\lambda}{d\lambda}=-2\,\lambda^{-3} \, \<\phi, H_0\phi>+k \lambda^{k-1}\, \<\phi, V\phi>=0.
$$

In particular, if $\phi$ is actually an eigenvector, then the
extremal is found for $\lambda=1$,
$$
2\, \<\phi, H_0\phi>=k\, \<\phi, V\phi>,
$$
and we reobtain in this way the virial theorem for eigenstates
of $H$.

\section{Final comments}

The standard theory of  virial theorem has been revisited from the perspective of  the theory of
Hamiltonian systems in symplectic manifolds.  This allows us to consider the classical counterpart of the usually called
 hypervirial theorems and clarify that the theory is not related in the general case with the group of scale transformations. In this 
 way we have found  room for dealing with systems whose configuration spaces are not linear spaces. 
 The geometric formalism is valid not only in the Hamiltonian approach (cotangent bundle with the canonical two-form) but also in the
  Lagrangian case (tangent bundle with a $L$-dependent symplectic form); we have shown the usefulness of some infinitesimal 
  point transformations that are not symmetries of the Lagrangian for establishing virial-like relations. The example of 
  position-dependent mass has been used to illustrate the theory and the particular example of a nonlinear oscillator is
  explicitly studied.

The use of  the modern symplectic approach  to quantum mechanics enables us to  prove in the second part of the paper that the virial theorem in quantum mechanics appears in full similarity with the analogous classical case.  We only need to take into account the point of view 
of symplectic geometry in infinite dimensional linear spaces. 
We hope this analogy will be helpful for dealing with various quantum cases.

\section*{Acknowledgments}

 This work was supported by the research projects FPA--2009--09638,
 MTM--2009--11154 (MEC, Madrid) and DGA-E24/1, DGA-E24/2 (DGA, Zaragoza).


\end{document}